\documentclass[aps,prb,twocolumn,superscriptaddress,showpacs]{revtex4-1}

\usepackage{graphicx}
\usepackage{calc}
\usepackage{bm}
\usepackage{color}
\usepackage{mathtools}

\begin{document}

\title{Surface spin polarization in the magnetic response of GeTe Rashba ferroelectric.}

\author{A.A. Avakyants}
\author{N.N. Orlova}
\author{A.V.~Timonina}
\author{N.N.~Kolesnikov}
\author{E.V.~Deviatov}
\affiliation{Institute of Solid State Physics of the Russian Academy of Sciences, Chernogolovka, Moscow District, 2 Academician Ossipyan str., 142432 Russia}

\date{\today}

\begin{abstract}
We experimentally investigate magnetization reversal curves  for a GeTe topological semimetal. In addition to the known lattice diamagnetic response, we observe  narrow magnetization loop in low fields, which should not be expected for non-magnetic GeTe. The hysteresis is unusual, so the saturation level is negative in positive fields, and the loop is passed clockwise, in contrast to standard ferromagnetic behavior. The experimental hysteresis curves can not be obtained from usual ferromagnetic ones by adding/subtracting of any linear dependence, or even by considering several interacting magnetic phases. The possibility of several phases is also eliminated by the remanence plots technique (Henkel or $\delta M$ plots). We explain our results as a direct consequence of the correlation between ferroelectricity and spin-polarized surface states in GeTe, similarly to magnetoelectric structures.  
\end{abstract}

\maketitle

\section{Introduction}

Recent renewal of interest to semimetals is mostly connected with topological effects.  Topological semimetals are conductors with gapless electronic excitations with band touching in some distinct points, which are protected by topology and symmetry~\cite{armitage}. Similarly to topological insulators~\cite{hasan} and quantum Hall systems~\cite{buttiker,ESreview}, topological semimetals have topologically protected  surface states. In Weyl semimetals (WSM) every band touching point splits  into two Weyl nodes with opposite chiralities due to the time reversal or inversion symmetries breaking. As a result, Fermi arc surface states connect projections of Weyl nodes on the surface Brillouin zone and these surface states inherit the chiral property of the Chern insulator edge states~\cite{armitage}.

Usually,  spin textures are known in magnetic materials as surface skyrmions~\cite{skyrm1,skyrm2,skyrm3,skyrm4,skyrm5,skyrm6,skyrm7,skyrm8,skyrm9} or  spin helix structures~\cite{shelix1,shelix2}. For the  magnetic WSMs (broken time reversal symmetry), the Fermi arc surface states were directly visualized in Co$_3$Sn$_2$S$_2$ by scanning tunneling spectroscopy~\cite{kagome_arcs}. Surface topological textures were visualized in some magnetic semimetals by STM, Lorenz electron microscopy, and magnetic force microscopy~\cite{CrGeTe,FGT_skyrmion1,FGT_skyrmion2}. Recent investigations show  topological protection  of skyrmion structures due to their origin from the spin-polarized topological  surface states~\cite{Araki}. 

However, spin textures due to the spin polarization of the Fermi arcs should also take place in nonmagnetic WSMs with broken inversion symmetry. Spin- and angle- resolved photoemission spectroscopy  technique has demonstrated spin-polarized surface Fermi arcs~\cite{das16,feng2016}. Spin-orbit interaction lifts the spin degeneracy of the surface states leading to their in-plane spin polarization, with strongly correlated and predominantly antiparallel spin textures in the neighboring Fermi arcs~\cite{Burkovetal2018}. As an example of nonmagnetic WSM,  spin polarization of the arcs reaches $80\%$, as it  has been discovered in TaAs~\cite{Xuetal2016}.   

Among nonmagnetic WSM materials, GeTe is of special interest~\cite{GeTespin-to-charge,GeTereview,GeTeour} due to the reported  giant Rashba splitting~\cite{GeTerashba,Morgenstern,GeTerashba1,GeTeour}. GeTe is predicted to be topological semimetal in  ferroelectric $\alpha$-phase~\cite{ortix,triple-point}.   
 Nonlinear Hall effect has been demonstrated in GeTe~\cite{GeTe2w}, which is the direct manifestation of finite Berry curvature in topological media~\cite{sodemann}.  
The direct measurement of the Rashba-split surface states of $\alpha$-GeTe(111) has been 
experimentally realized thanks to K doping~\cite{GeTesurfStates}. It has been shown that the 
surface states are not the result of band bending and that they are decoupled from the bulk states. The giant Rashba splitting of the surface states of $\alpha$-GeTe is largely arising from the inversion symmetry breaking in the bulk~\cite{GeTesurfStates}. Also, direct correlation between ferroelectricity and spin textures was demonstrated in GeTe~\cite{spin text}. 

Surface spin polarization have been directly demonstrated in a magnetic response of topological semimetals with broken time-reversal symmetry~\cite{cosimag,cosnsmag}.
Thus, one  can expect a  complicated response of a topological semimetal GeTe on the external magnetic field due to the spin polarization of topological surface states. 

Here, we experimentally investigate magnetization reversal curves  for a GeTe topological semimetal. In addition to the known lattice diamagnetic response, we observe  narrow magnetization loop in low fields, which should not be expected for non-magnetic GeTe. We explain our results as a direct consequence of the correlation between ferroelectricity and spin-polarized surface states in GeTe, similarly to magnetoelectric structures.

\section{Samples and technique}

\begin{figure}
\includegraphics[width=1\columnwidth]{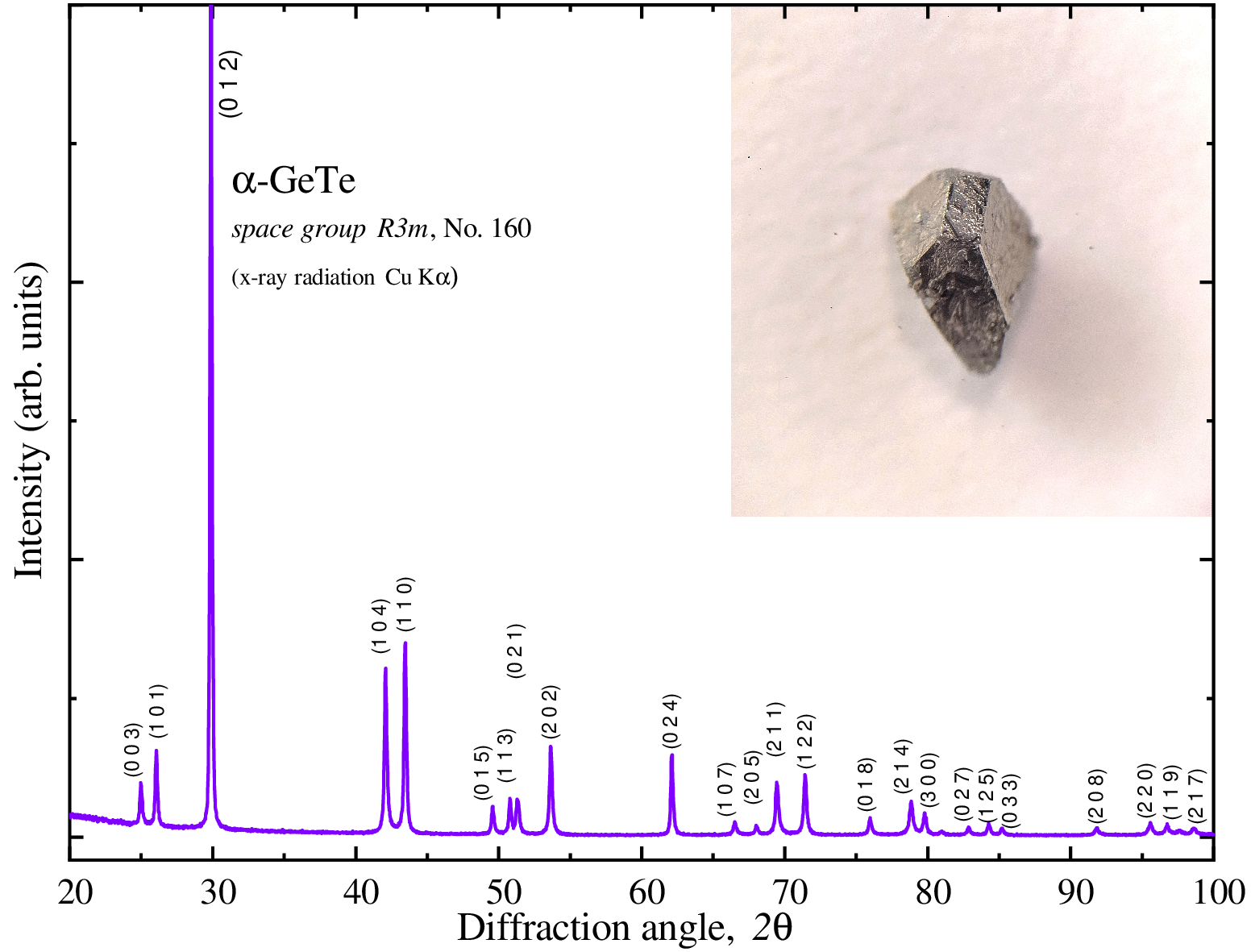}
\caption{(Color online) (a) The X-ray powder diffraction  pattern (Cu K$\alpha$ radiation), which is obtained for the crushed GeTe single crystal. The single-phase  $\alpha$-GeTe is confirmed with the space group R3m (No.160).
The  inset shows optical image of the crystal.
  }
\label{sample}
\end{figure}

GeTe single crystals were grown by physical vapor transport in the evacuated silica ampule. The initial GeTe load was synthesized by direct reaction of the high-purity (99.9999\%) elements in vacuum. For the  crystals growth, the obtained GeTe serves as a source of vapors: it was melted and kept at 770-780$^\circ$ C for 24 h. Afterward, the source was cooled down to 350$^\circ$ C at the 7.5 deg/h rate.  GeTe crystals grew during this process on the cold  ampule walls  above the source. 

The GeTe composition is verified by energy-dispersive X-ray spectroscopy. The powder X-ray diffraction analysis confirms single-phase GeTe, see Fig.~\ref{sample} (a),  the known structure model~\cite{GeTerashba} is also refined with single crystal X-ray diffraction measurements. 
Ferroelectric polarization and Rashba splitting are defined by the non-centrosymmetric distorted rhombohedral structure ($\alpha-GeTe$) with space group R3m (No. 160)~\cite{GeTerashba}. 
For our GeTe single crystals, giant Rashba splitting~\cite{GeTerashba} has been confirmed in capacitance measurements~\cite{GeTeour}.

\begin{figure}
\includegraphics[width=\columnwidth]{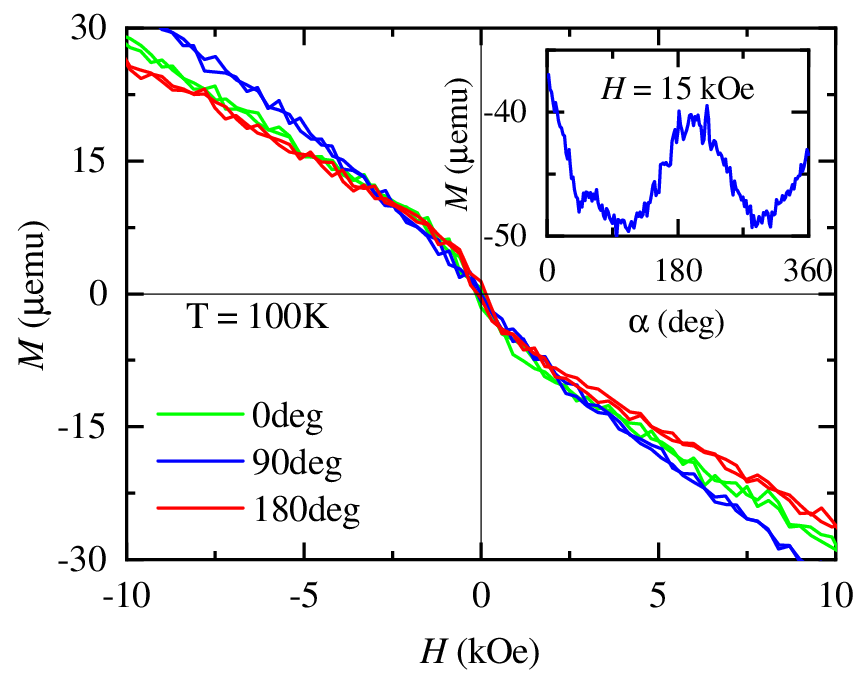}
\caption{(Color online) Magnetization curves at 100~K temperature for the 6.69~mg GeTe flake.  The overall $M(H)$ behavior shows diamagnetic response~\cite{GeTe_diamag}, which is accompanied by the clearly visible kink in $M(H)$ dependence at low fields, within $\pm1$~kOe interval.  The $M(H)$ curves are shown for several angles $\alpha$ between the sample holder and magnetic field. Inset shows   $M(\alpha)$ dependence: we observe about 20\%  modulation of the $M(\alpha)$ with 180$^\circ$ periodicity in the magnetic field 15~kOe. This shallow  angle dependence seems to originate from the shape of the exfoliated flake with well-developed cleaved surface.   
 }
\label{wide hyster}
\end{figure}

To investigate magnetic properties, we use Lake Shore Cryotronics 8604 VSM magnetometer, equipped with nitrogen flow cryostat.  The topological semimetals are essentially three-dimensional objects~\cite{armitage}, so we have to select  relatively thick (above 0.5~$\mu$m) mechanically exfoliated GeTe flakes. A small (0.82--9.54~mg)  flake is mounted to the sample holder by  low temperature grease, which has been tested to have a negligible magnetic response. 

We investigate sample magnetization by standard method of the magnetic field gradual sweeping between two opposite field values to obtain magnetization loops.  Also, the remanence plots technique (i.e., Henkel or $\delta M$ plots) is routinely used to evaluate interactions between nanoparticles or grains~\cite{Goey,Kelly,Henkel}.

The technique is based on the comparison of the isothermal remanent magnetization curve (IRM, $M_r$), and the DC demagnetization remanence curve (DCD, $M_d$). The IRM curve is obtained on an initially demagnetized sample by applying a positive magnetic field. The DCD curve is measured by first saturating the sample and then measuring the remanence magnetization after application of progressively larger fields of opposite direction.  For a system of noninteracting single-domain particles with uniaxial anisotropy, the IRM and DCD are related to each other via the Wohlfarth equation
$$
M_d(H) = M_{rs}-2M_r(H)
$$
where $M_{rs}$ is the saturation remanence and $H$ is the applied magnetic ﬁeld. 

The $\delta M$ or  Henkel plot  is a direct measure of the deviation from the linearity:
$$
\delta M ( H ) = M_d(H)-[M_{rs}-2M_r(H)]
$$
Interparticle interactions are detected through the appearance of a negative dip
(demagnetizing interactions, typically dipolar one) or a positive peak (magnetizing, usually exchange, interactions) in the $\delta M$ plots, whereas $\delta M = 0$ has generally been taken as an indication of the absence of interactions~\cite{Kelly,Bender}. In other words, positive and negative $\delta M$  contributions indicate  more than one phase~\cite{Sanchez}.

\section{Experimental results}

Fig.~\ref{wide hyster} shows magnetization loops at 100~K temperature for the 6.69~mg   GeTe flake.
The overall $M(H)$ behavior shows the  diamagnetic response,  which is known for the bulk GeTe mostly due to the lattice~\cite{GeTe_diamag}. From the linear diamagnetic dependence we can estimate the slope as $ -3\cdot 10^{-6}$ emu/cm$^3$,  this estimation well corresponds to the reported GeTe volume susceptibility~\cite{GeTe_diamag}. The $M(H)$ curves show some angle dependence, as it is confirmed  by the direct $M(\alpha)$ measurement in the inset to Fig.~\ref{wide hyster}: we observe about 20\%  modulation of the $M(\alpha)$ with 180$^\circ$ periodicity in the magnetic field 15~kOe. This shallow  angle dependence seems to originate from the shape of the exfoliated flake with well-developed cleaved surface.

\begin{figure}
\includegraphics[width=\columnwidth]{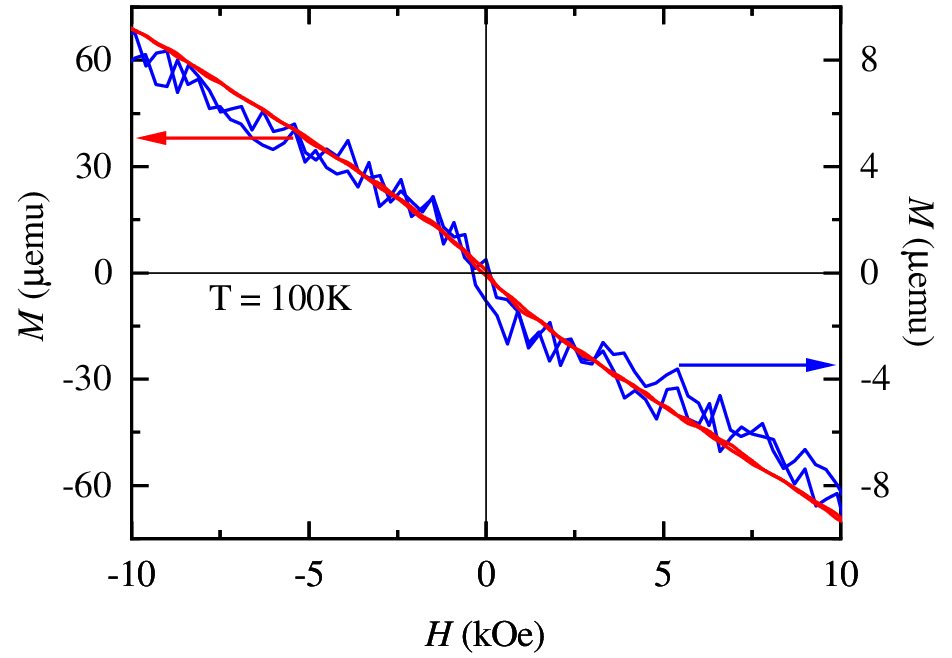}
\caption{(Color online) Magnetization loops at 100~K temperature for two 9.54~mg and 0.82~mg GeTe flakes, see the left and the right axes, respectively. The diamagnetic slope  scales with the sample mass. The linear curves with low-field kink are qualitatively similar to the behavior in  Fig.~\ref{wide hyster}.  The low-field kink is better seen for the smaller sample, so even low-field hysteresis can be seen for the smallest 0.82~mg GeTe flake.
 }
\label{diff samples}
\end{figure}

The most striking experimental result is the clearly visible kink in $M(H)$ dependence at low fields, within $\pm1$~kOe interval,  see Fig.~\ref{wide hyster}. The kink can be seen for any sample orientation. Diamagnetic response with low-field kink  can be qualitatively reproduced for different GeTe flakes. For example, Fig.~\ref{diff samples} shows $M(H)$ curves for 9.54~mg and 0.82~mg samples, see the left and the right axes, respectively. We should conclude, that the standard linear diamagnetic response is accompanied by narrow magnetization loop, which should not be the case for the diamagnetic GeTe.

The diamagnetic slope scales with the sample mass, as it should be expected for the lattice-induced response: it is increased in approximately two times for the 1.5 mass increase, cp. Fig.~\ref{wide hyster} and Fig.~\ref{diff samples}, left axis; also, the slopes differ in 8 times for two flakes in Fig.~\ref{diff samples}, which is near the sample mass ratio 11.6. The $\approx 25\%$ discrepancy can to be ascribed to the different shape of the exfoliated flakes, due to the arbitrary orientation of the cleaved surfaces (cp. with the 20\%  modulation of the $M(\alpha)$  in the inset to Fig.~\ref{wide hyster}).

\begin{figure}
\includegraphics[width=\columnwidth]{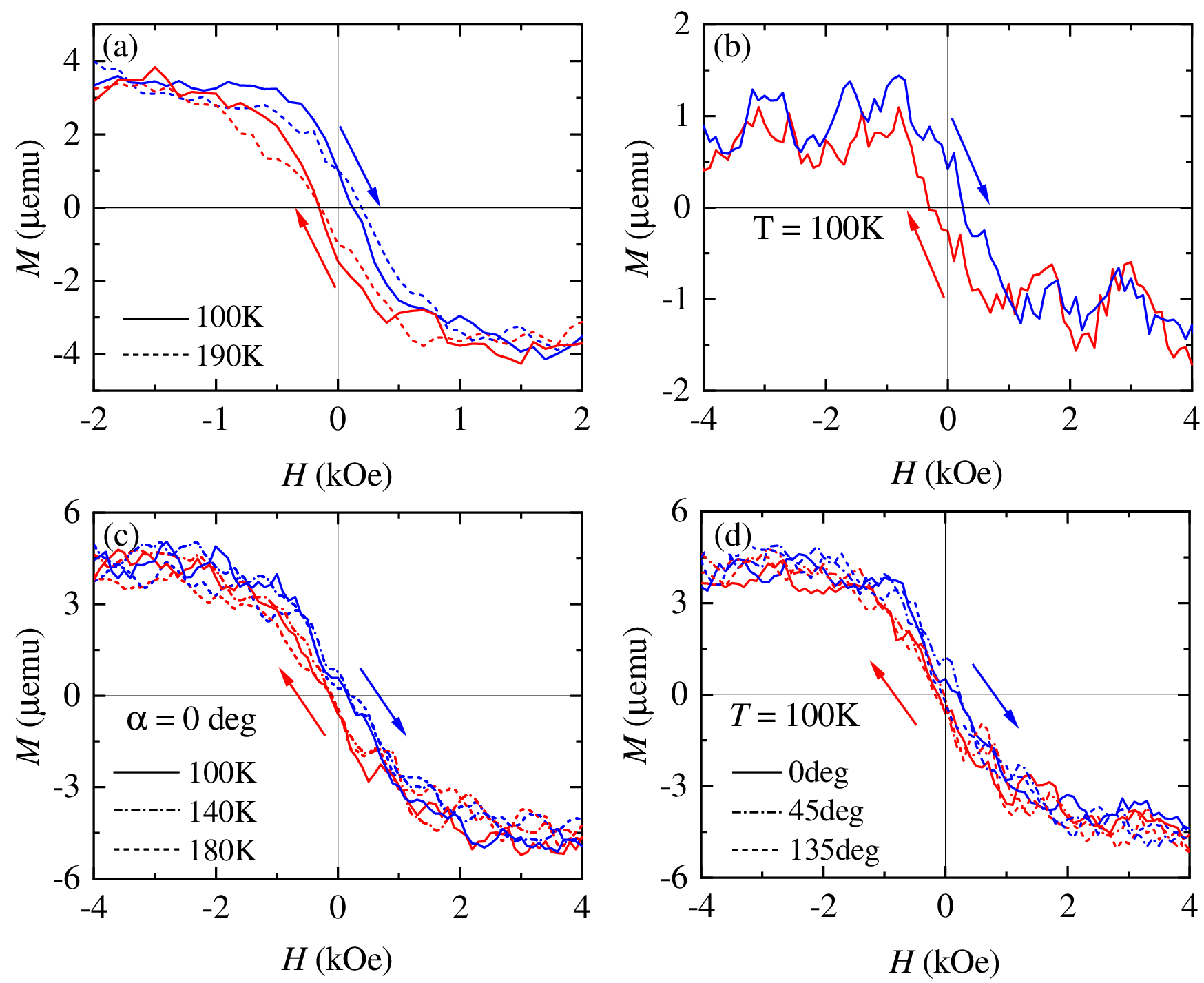}
\caption{(Color online) The low-field hysteresis region for all three samples. For every field sweep direction, we use curve averaging (8 curves) to increase the signal/noise ratio. The diamagnetic slope is subtracted from the averaged curves to highlight the nonlinear low-field behavior. (a) Curves for the 6.69~mg GeTe flake at two temperatures, 100~K (solid) and 190~K (dash). (b) $M(H)$ curves for the smallest, 0.82~mg   GeTe flake, at 100~K. (c-d) Hysteresis for the 9.54~mg flake at different temperatures (c) and sample orientation (d). For every sample, the saturation level is negative in positive fields, and the loop is passed clockwise, in contrast to usual ferromagnetic hysteresis.  
  }
\label{narrow hyster}
\end{figure}

The low-field kink is better seen for the smaller samples, so even low-field hysteresis can be seen for the smallest 0.82~mg GeTe flake in Fig.~\ref{diff samples}. The hysteresis is shown in detail in Fig.~\ref{narrow hyster} for all three samples. We use curve averaging (8 curves) to increase the signal/noise ratio. The linear diamagnetic slope is subtracted from the averaged curves to highlight the nonlinear low-field behavior. 

First of all, all three samples show clear low-field hysteresis in Fig.~\ref{narrow hyster}. To our surprise, the  saturation level is negative in positive fields, and the loop is passed clockwise, in contrast to usual ferromagnetic hysteresis. We wish to note, that the experimental curves in Fig.~\ref{narrow hyster} can not be continuously transformed to the standard ferromagnetic one (the saturation level is positive in positive fields) by adding/subtracting of any linear dependence: the linear diamagnetic curve, been subtracted from the magnetization loop, can not invert the saturation levels around the zero field. This even qualitatively excludes any possible contribution from the sample holder, nitrogen atmosphere, and so on. 

The saturation level value monotonically depends on the sample mass in Fig.~\ref{narrow hyster} (a) and (c): it is increased from 3~$\mu$emu in (a), for the 6.69~mg   GeTe flake, to 4.5~$\mu$emu in (c), for the 9.54~mg   one. It is below 1~$\mu$emu in (b), for the smallest, 0.82~mg   GeTe flake, but the signal is noisy here. The hysteresis width at zero $M$ level (coercitivity) is different for all three samples, but no reasonable dependence can be seen in Fig.~\ref{narrow hyster} (a-c). The hysteresis is not sensitive to temperature below 200~K, as it is shown in Fig.~\ref{narrow hyster} (a) and (c). After subtracting the diamagnetic slope, the low-field hysteresis is not sensitive to the field orientation within our accuracy, see Fig.~\ref{narrow hyster} (d). 

\begin{figure}
\includegraphics[width=\columnwidth]{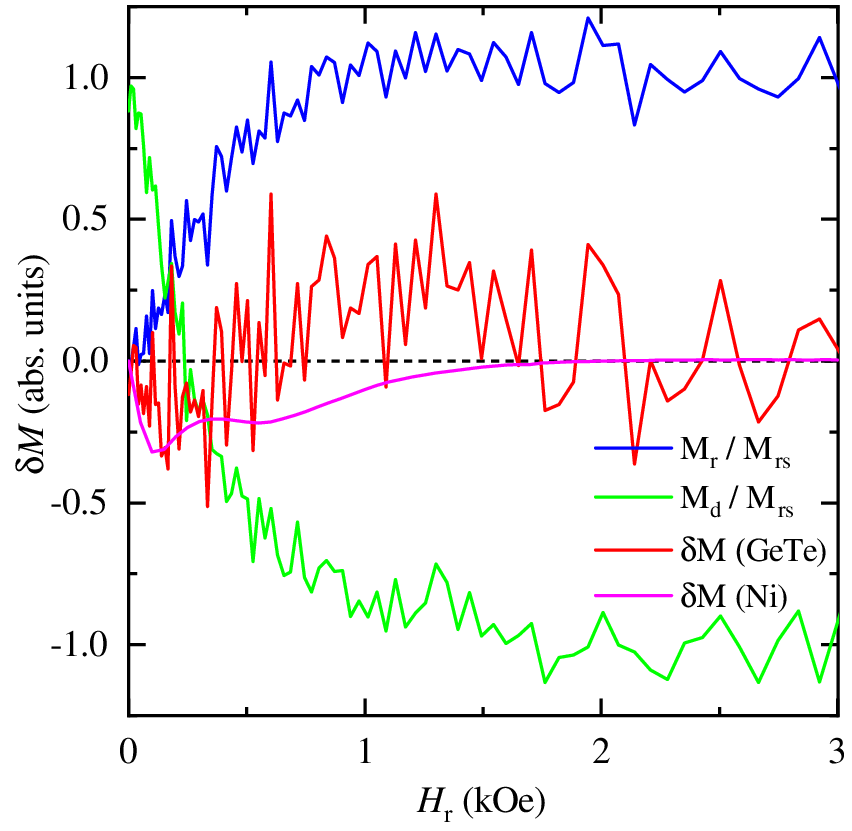}
\caption{(Color online) Henkel or remanence $\delta M$ plot for 6.69~mg GeTe flake at 100~K temperature. The initial $M_r(H)$,$M_d(H)$ curves (blue and green ones, respectively), give the $\delta M(H)$  (red) curve. Within the accuracy of experiment, $\delta M(H)$ behavior can not definitely confirm several magnetic phases for our GeTe flakes. For comparison, a reference fully negative $\delta M$ curve (magenta) is shown, as obtained for the standard  nickel sample. 
 }
\label{delta M}
\end{figure}

In principle, the hysteresis in  Fig.~\ref{narrow hyster} could be connected with several (at least two) interacting magnetic phases. This can be verified by the remanence plots technique (i.e., Henkel or $\delta M$ plots), see Fig.~\ref{delta M}. Positive and negative $\delta M$  contributions indicate  more than one phase~\cite{Sanchez}, whereas $\delta M = 0$ has generally been taken as an indication of the absence of interphase interactions~\cite{Kelly,Bender}.

Fig.~\ref{delta M} shows both the initial $M_r(H)$,$M_d(H)$ curves and the calculated $\delta M(H)$. $\delta M(H)$ for GeTe varies around zero within the accuracy of experiment. Fig.~\ref{delta M} also shows a reference  $\delta M$ curve, obtained for the standard  nickel sample. In the latter case, negative  $\delta M$ dip corresponds to the dipolar interaction between domains. Thus,   the remanence plots technique can not definitely confirm several magnetic phases for our GeTe flakes.

\section{Discussion} \label{disc}

As a result, we observe that the lattice diamagnetic response is accompanied by the low-field hysteresis loop in GeTe. In contrast to usual ferromagnetic hysteresis, the  saturation level is negative in positive fields so the loop is passed clockwise.

This hysteresis can not be obtained from standard ferromagnetic one even by considering several interacting magnetic phases. In the latter case the bias field can move the magnetization switching even across the zero, so the magnetization loop is passed clockwise~\cite{cosnsmag} (in contrast to single-phase ferromagnet with usual counterclockwise hysteresis). However, the bias field can not invert the saturation levels.  Also, the remanence plots technique (i.e., Henkel or $\delta M$ plots) does not confirm several magnetic phases for our GeTe flakes.   

Even qualitatively, one can not also find two interacting magnetic phases in GeTe.  One of the phases could be the spin textures from the surface states~\cite{GeTesurfStates} in $\alpha$-GeTe(111). As it has been shown before, spin-polarized surface states can, in principle, give significant contribution into the overall magnetic response~\cite{cosnsmag,cosimag}. The second phase is the ferromagnetic bulk in magnetic topological semimetals~\cite{cosnsmag}, but it can not be expected for the diamagnetic GeTe. 

Also, GeTe composition is verified by energy-dispersive X-ray spectroscopy and the powder X-ray diffraction analysis. The obtained volume susceptibility $ -3\cdot 10^{-6}$ well corresponds to the known values~\cite{GeTe_diamag}. Thus, there is  no magnetic  impurities in our GeTe crystals. 

On the other hand, correlation between the Rashba-split surface states~\cite{GeTesurfStates}  and ferroelectricity in $\alpha$-GeTe can be responsible for the observed hysteresis. Both the giant Rashba splitting of the surface states and bulk ferroelectricity are largely arising from the inversion symmetry breaking~\cite{GeTesurfStates}. For our GeTe single crystals, giant Rashba splitting~\cite{GeTerashba} and bulk ferroelectricity have been confirmed in capacitance measurements~\cite{GeTeour}.  Direct correlation between ferroelectricity and spin textures was demonstrated in GeTe~\cite{spin text}. Thus, GeTe single crystal can be considered as magnetoelectric heterostructure~\cite{magnetoelecric1,magnetoelecric2} or multiferroic system~\cite{MForigin}. 

In magnetoelectrics, due to the coupling among the different degrees of freedom (ferroelectricity, ferromagnetism, or ferroelasticity) leading to these ordered states, the order parameter of one state can be controlled by tuning parameters different from their conjugate variable~\cite{MForigin}.  In the conditions of our experiment, variation of the magnetic field leads to appearance of the electric field due to the magnetoelectric coupling~\cite{indusi}. Electric field affects spin textures in GeTe~\cite{spin text,magnetoelecric2}, which, subsequently, affects magnetization response~\cite{cosnsmag,cosimag}. As a result, the net magnetization follows the external field with some delay, producing the unusual hysteresis around zero field. This effect can be clearly seen only for significant diamagnetic response in GeTe, where it can not be mixed with usual ferromagnetic loop. Thus, unusual low-field hysteresis  is a direct consequence of correlation between ferroelectricity and spin-polarized surface states in GeTe.

\section{Conclusion}
As a conclusion, we experimentally investigate magnetization reversal curves  for a GeTe topological semimetal. In addition to the known lattice diamagnetic response, we observe  narrow magnetization loop in low fields, which should not be expected for non-magnetic GeTe. The hysteresis is unusual, so the saturation level is negative in positive fields, and the loop is passed clockwise, in contrast to standard ferromagnetic behavior. The experimental hysteresis curves can not be obtained from usual ferromagnetic ones by adding/subtracting of any linear dependence, or even by considering several interacting magnetic phases. The possibility of several phases is also eliminated by the remanence plots technique (Henkel or $\delta M$ plots). We explain our results as a direct consequence of the correlation between ferroelectricity and spin-polarized surface states in GeTe, similarly to magnetoelectric structures.

\acknowledgments

We wish to thank S.S~Khasanov for X-ray sample characterization.  We gratefully acknowledge financial support  by the  Russian Science Foundation, project RSF-23-22-00142, https://rscf.ru/project/23-22-00142/.

\end{document}